\documentstyle[12pt,epsfig]{article}
\input epsf
\newcommand{\be}{\begin{equation}}
\newcommand{\ee}{\end{equation}}

\newcommand{\gsim}{\:\raisebox{.25ex}{$>$}\hspace*{-.75em}
      \raisebox{-.93ex}{$\sim$}\:}
\newcommand{\lsim}{\:\raisebox{.25ex}{$<$}\hspace*{-.75em}
      \raisebox{-.93ex}{$\sim$}\:}

\begin{document}

\begin{titlepage}
\centerline{\large\bf INSTITUTE OF THEORETICAL AND EXPERIMENTAL PHYSICS}
\flushright{95/41}

\vspace{8cm}
\centerline{\large\bf A.~Dobrovolskaya ~and~ V.~Novikov }
\vspace{3cm}

\centerline{\large\bf FACTORIZATION ~METHOD }
\vspace{5mm}
\centerline{\large\bf IN ~THE ~SPONTANEOUSLY ~BROKEN ~GAUGE ~THEORIES }
\vspace{7cm}

\centerline{\large\bf Moscow ~~---~~ 1995}

\end{titlepage}

\newpage
$~~$
\vspace{60mm}
\centerline{\large\bf A b s t r a c t}
\vspace{5mm}

     We consider the factorization method in the spontaneously broken
gauge theories such  as the electroweak theory.

     Using the of\/f-shell $W$-boson density matrix formalism
we demonstrate that the
factorization conditions are completely under control.

     The main point of this paper is the presence of the
"interference" or "crossed" terms in $WW$--scattering process
which exhibits itself in
the dependence on the relative azimuthal angle
between two scattering planes.

     To illustrate our general consideration we  use
well-known example --- a single Higgs boson production. The
origine of the quantitative failure of the $WW$--ef\/fective method, as
originally propossed, is given and it is shown how to use it
correctly.

\newpage
%%%%%%%%%%%%%%%%%%%%%%%%%%%%%%%%%%%%%%%%%%%%%%%%%%%%%%%%%%%%%%%%%%%%%%%%%%

\underline{\large\bf I. ~I n t r o d u c t i o n}
\vspace{3mm}

%%%%%%%%%%%%%%%%%%%%%%%%%%%%%%%%%%%%%%%%%%%%%%%%%%%%%%%%%%%%%%%%%%%%%%%%%

In order to search for a new physics at future $e^+e^-$--colliders or at
LHC we must well understand the fusion mechanism especially via
vector $W, Z$ bosons.

Our paper is devoted  to the applicability of the effective $W$-boson
approximation (EWA)-method. This approximate method was proposed for
the calculation of $WW(ZZ)$--fusion processes, for which there were no
hopes to find exact analytical answers in general case.

Originally [1, 2] only longitudinally polarized $W$-boson beams
were taken into account by EWA. It was found that the result of calculation
of the cross-section for Higgs-boson production via $WW$-fusion by EWA
(in the region of $W$-boson energy $E_W \gsim m_H > m_W$)
is much~ \underline{larger}
{}~than the exact total cross-section [3]. The contribution of the
transversaly polarized $W$-beams to this process only increases the
discrepancy between this classical term and the exact total cross-section.

A lot of ef\/forts were spent to resolve this discrepancy between exact
computer calculations and the EWA--method. But the results were not
very succesful -- all the corrections were found to be small and the
EWA--method was announced to be correct only for heavy Higgs--boson
($M_H \gg m_W$) and only in high energy limit $E_W \gg M_H$ [4, 5].

In this paper we suggest a resolution of this problem based on the
advanced version of EWA--method developed in our paper [6]. Actually
this method is well known in gamma-gamma physics [7].

The main point is that the beams of~ \underline{virtual} ~$W$-boson are
described not only by probabilities to find $W$-boson with
longitudinal and transversal polarization but by density matrix (as
it should be in quantum mechanics).
We consider the~ \underline{nondiagonal} ~terms in the
virtual  $W$--boson density matrix, which turn out to be
of the same order of magnitude as the diagonal ones (corresponding to the
longitudinal and transversal polarizations).
If $e^+e^-$ (or $q \bar q$) lab. frame and $W^+W^- (ZZ)$ s.m.f.
are collinear the contribution from these nondiagonal elements of
density matrix to the total cross section is equal to zero.
This is the reason why nondiagonal terms were not considered preveously.
But $e^+e^-$ (or $q \bar q$ ) lab.
frame and $W^+ W^- (ZZ)$ c.m.f.  are not exactly
collinear and  the integration
over azimuthal angle in the lab. frame does not cancel the
contribution from non-diagonal density matrix terms. This means that
two $W$ boson beams can not be considered as an independent, i.e.
factorization property is violated in this kinematical region. We
have found that these interference terms, which having a negative
sing and a large absolute value, explain the discrepancy between EWA
and exact computer calculation. To restore factorization (and EWA
method) we have to conf\/ine kinematical region to smaller scattering
angles. We demonstrate that the EWA--method correctly applied yields
appropriate lower bounds for the total cross-section for any energy
of colliding particles and any (small or large) mass of Higgs-boson.
The contribution of each element into the cross-section depends on
the dynamics of the $WW \to "X"$ subprocess (where $"X"$ can be a
Higgs boson(s), a $W, Z$ -- boson pair or a lepton pair).  For
example for the $WW \to H$ process (described as an example in this
article) the contribution of longitudinal and spin-flip interference
terms are the most important ones and the transversely polarized
term's contribution is negligible (contrary to [4] P.Johnson et al.).

The paper is organized in the following way: the helicity density matrix
for  virtual $W$ and  $Z$  bosons are defined in Section
II.  To illustrate our general consideration we use simple and
well-known example -- a single Higgs-boson production (see
Section III).  In Section IV we have discussed the exact numerical
calculations of the contribution of each term in the helicity decomposition
into the total cross-section.
In Section V we define the kinematical region for the application
of the ef\/fective $W$-boson method. The results are discussed in the
Conclusions.

Some results presented here, have been already published in our recent
preprint [8]. In [6] we have used this approach for two Higgs boson
production via $WW$--fusion.
\vspace{5mm}

%%%%%%%%%%%%%%%%%%%%%%%%%%%%%%%%%%%%%%%%%%%%%%%%%%%%%%%%%%%%%%%%%%%%%%%%%%

\underline{\large\bf  II. ~$W$--boson density matrix}
\vspace{3mm}

%%%%%%%%%%%%%%%%%%%%%%%%%%%%%%%%%%%%%%%%%%%%%%%%%%%%%%%%%%%%%%%%%%%%%%%%%

To describe the polarization properties of the virtual $W$-boson we
introduce the basis of virtual $W$-boson helicity states. It could be
done in the center of mass system (c.m.s.) of $W^+$ and $W^-$
$$
  q_1 = (w_1, 0, 0, q)~; ~~~~~~~~~q_2 = (w_2, 0, 0, -q)~.
$$
Consider the standard set of orthonormal four-vectors orthogonal to the
momentum $q_1$:
% 1
\be
\begin{array}{lll}
  e^{(1)}_{\mu} (+1) = \frac{1}{\sqrt 2} (0, -1, -i, 0) ~, \\ \nonumber
      e^{(1)}_{\mu} (0) = \frac{1}{\sqrt{-q^2_1}} (q, 0, 0, w_1)
  \stackrel{def}{\equiv} - i Q_1 ~, \\
      e^{(1)}_{\mu} (-1) = \frac{1}{\sqrt 2} (0, 1, -i, 0)
\end{array}
\ee
and the analogous set for the second $W$-boson
$$
  e^{(2)}_{\mu} (\pm 1) = e^{(1)}_{\mu} (\mp 1) ~,
$$
% 2
\be
  e^{(2)}_{\mu} (0) = \frac{i}{\sqrt{-q^2_2}} (-q, 0, 0, w)
  \stackrel{def}{\equiv}  i Q_2 ~,
\ee
$$
  e^{(2)}_{\mu} \cdot q_{2\mu} = 0 ~.
$$
These four-vectors obviously represent the $\pm 1$ and $0$ helicity
states of virtual $W$-bosons in their c.m.s. and form a complete orthonormal
basis for the subspace orthogonal to $q_{1\mu}$ and $q_{2\mu}$
respectively:
% 3
\be
   e^{(1) \star}_{\mu} (a) e^{(1)}_{\mu} (b)  = (- 1)^a \delta_{a,b} ~,
   ~~~~~~~~~~a, ~b = \pm 1, ~0 ~,
\ee
% 4
\be
   \sum\limits_a e^{(1) \star}_{\mu} (a) e^{(1)}_{\nu} (a)  =
   g_{\mu\nu} - \frac{q_{1\mu} q_{1\nu} }{q^2_1} ~.
\ee
Any vector or matrix orthogonal to $q_{1\mu}$ could be expanded over
this basis. Further we shall neglect the masses of light quarks and
leptons. In this approximation
%5
\be
   \rho^{(1)}_{\mu\nu} q_{1\mu} ~=~ \rho^{(2)}_{\mu\nu} q_{2\mu} ~=~  0
\ee
and as a result
% 6
\be
   \rho^{(1)}_{\mu\nu} = \sum\limits_{a, b}
   e^{(1) \star}_{\mu}  e^{(1)}_{\nu} (b) \rho^{(1)}_{a, b} ~,
   ~~~~~~~a ~= \pm 1, ~0 ~,
\ee
% 7
\be
   \rho^{(1)}_{ab} = (- 1)^{a + b}
   e^{(1)}_{\mu}  (a) [ e^{(1)}_{\nu} (b) ]^{\star} \rho^{(1)}_{\mu, \nu} ~.
\ee
Here $\rho_{a b}$ is the density matrix in the helicity representation:
% 8
\be
   \rho^{(i)}_{a b} = \left\{
\begin{array}{ccc}
   \rho^{(i)}_{+ +}  ~~\rho^{(i)}_{+ 0}   ~~\rho^{(i)}_{+ -}   \\[2mm]
   \rho^{(i)}_{0 +}  ~~\rho^{(i)}_{0 0}   ~~\rho^{(i)}_{0 -}   \\[2mm]
   \rho^{(i)}_{- +}  ~~\rho^{(i)}_{- 0}   ~~\rho^{(i)}_{- -}
\end{array}
    \right\} ~, ~~~~~i = 1, ~ 2 ~.
\ee

Detailed expressions for Lorentz invariant density matrix elements
$\rho^{(i)}_{a b}$ are presented in the Appendix I and II. This density
matrix is nondiagonal. It means that the system is strongly polarized
and quantum mechanical interference ef\/fects are large. The
diagonal components $\rho^{(i)}_{+ +},~ \rho^{(i)}_{- -}$ and
$\rho^{(i)}_{0 0}$ represent the fraction of longitudinal and transverse
$W$--bosons inside the fermion (electron) (unnormalyzed classical
probabilities). The nondiagonal components $\rho^{(i)}_{+ 0},~
\rho^{(i)}_{0 -}$ and $\rho^{(i)}_{+ -}$ correspond to the spin-f\/lip
transitions of $W$--boson and depend on the azimuthal angle
$\tilde{\varphi}_1$ in the $W^+W^-$ c.m.s. For the relative azimuthal
angle between the scattering planes of the colliding particles
$\Delta \tilde{\varphi} = \tilde{\varphi}_1 - \tilde{\varphi}_2$,
we have:
% 9
\be
   \cos \Delta \tilde{\varphi} = - \frac{(p_{1 \perp} p_{2 \perp})}
    {\sqrt{p^2_{1 \perp} \cdot p^2_{2 \perp}}} ~,
\ee
where ~$p_{1 \perp} ~~(p_{2 \perp})$ ~is ~the ~perpendicular ~component
{}~of ~the ~electron ~momentum.
    \footnote{\normalsize  Note, that the $W$-density matrix is def\/ined
    only by the fermion-$W$-fermion vertex and have no dependence on the
    Higgs-boson models.  }

$W$-bosons are produced by the left fermion component only, that is why
the $W$-boson density matrice elements are dif\/ferent from the photon
density matrice ones [7], as opposed to photons, which produced by the
left and right fermions with the same weight, and to $Z$-bosons ones
(because $Z$--boson is
produced by left and right fermion components, but with the dif\/ferent
weights). These tiny ef\/fects are interesting to study at polarized
$e^+e^-$--beams. We shall discuss them in the next publication.
\vspace{5mm}

%%%%%%%%%%%%%%%%%%%%%%%%%%%%%%%%%%%%%%%%%%%%%%%%%%%%%%%%%%%%%%%%%%%%%%%%

\underline{\large\bf III. ~Exact calculation of Higgs-boson production}
\vspace{3mm}
%%%%%%%%%%%%%%%%%%%%%%%%%%%%%%%%%%%%%%%%%%%%%%%%%%%%%%%%%%%%%%%%%%%%%%%%%

The process of Higgs-boson production via $W$-boson fusion in
$e^+e^- $  (or $q \bar q)$  collisions is shown in (Fig. 1). Its
cross-section can be written in the form:
% (10)
\be
  d \sigma = \left( \frac{\pi \alpha_{w}}{2} \right)^2~ \cdot
  \frac{1}{4 \sqrt{(p_1 p_2)^2 - p^2_1 p^2_2}}
  ~\frac{q^2_1}{(q^2_1  - m^2_w)^2}
  ~\frac{q^2_2}{(q^2_2  - m^2_w)^2} \cdot
\ee
$$
  \left[ 4 \rho^{(1)}_{\mu \nu}  \rho^{(2)}_{\alpha \beta} \right]
   M^{\star}_{\mu \alpha}  M_{\nu \beta} (2 \pi)^4  \delta^4
   (q_1 + q_2 - k) \frac{d^3 \vec k}{(2 \pi)^3 2 E_k}
   \frac{d^3 \vec p~'_1}{(2 \pi)^3 2 E'_1} \cdot
   ~\frac{d^3 \vec p~'_2}{(2 \pi)^3 2 E'_2}  ~,
$$
where $\alpha_w = \frac{\alpha}{\sin^2 \theta_w}$
($\alpha$ -- is the fine structure constant and $\theta_w$ is a Weinberg
angle);
$M_{\mu \alpha}$--denotes the amplitude for Higgs-boson production in the
virtual $W$-boson collision, in the Standard model:
% 11
\be
   M_{\mu \alpha } ~ = ~ g m_w \cdot g_{\mu \alpha}  ~.
\ee
Note, that the tensor structure -- $g_{\mu \alpha}$ of this vertex is
f\/ixed by the unitarity and is thus the same in all models, except
for some scaling coef\/ficients [9]. The tensors $\rho^{(1)}_{\mu \nu}$
and $\rho^{(2)}_{\alpha \beta}$ are def\/ined by the currents of the
colliding particles (electrons):
% 12
\be
   \rho^{(1)}_{\mu \nu} ~ = ~ - \frac{1}{2q^2_1}
    Tr \left[ (\hat p_1 + m_e) \gamma_{\mu} (\hat p'_1 + m_e) \gamma_{\nu}
    (1 + \gamma_5) \right]  =
\ee
$$
   = - \left( g_{\mu \nu}  - \frac{q_{1 \mu} q_{1 \nu}}{q^2_1}  \right)
     - \frac{(2p_1 - q_1)_{\mu}(2p_1 - q_1)_{\nu}}{q^2_1}  +
     2 i \epsilon_{\mu \nu \alpha \beta} p_{1 \alpha}
      q_{1 \beta} + g_{\mu \nu} \frac{m^2_e}{q^2_1}
$$
and represent spin density matrix for ef\/fective $W$-bosons.

     If we substitute (11, 12) in formula (10) we shall obtain the known
[10, 11, 13] expression for the matrix element squared:
% 13
\be
   \left[ \rho^{(1)}_{\mu \nu} \rho^{(2)}_{\alpha \beta} \right]
    M^{\star}_{\mu \alpha} M_{\nu \beta}  =
    \frac{16}{q^2_1 q^2_2} (p_1 p_2) (p'_1 p'_2)
\ee
where $p_1,  p_2 ~(p'_1, p'_2)$ are the four-momenta of the incident
(final) electrons (quarks) (Fig.1), we have verif\/ied that the
exact formula (10) with (13) gives numerically the same result [12]
as the previous calculations [2, 10, 11]. (One integration
with $\delta$-function has been done analytically, see Appendix III.)
Below, we shall discuss the contribution of the dif\/ferent
$W$-boson polarization states in the total cross-section.
\vspace{5mm}

%%%%%%%%%%%%%%%%%%%%%%%%%%%%%%%%%%%%%%%%%%%%%%%%%%%%%%%%%%%%%%%%%%%%%%%%

\underline{\large\bf  IV. ~Exact calculation of the input of dif\/ferent}
\underline{\large\bf      $W$-boson polarization states
                          into the total cross-section}
\vspace{3mm}
%%%%%%%%%%%%%%%%%%%%%%%%%%%%%%%%%%%%%%%%%%%%%%%%%%%%%%%%%%%%%%%%%%%%%%%%%

Using the helicity representation for $\rho^{(1)}_{\mu \nu}$ and
$\rho^{(2)}_{\mu \nu}$  (6) - (8) one can rewrite the total cross-section
(10) in the following way [6]
% 14
$$
   d \sigma = \frac{1}{4(p_1 p_2)} \left( \frac{\pi \alpha_w}{2} \right)^2
   \frac{q^2_1}{(q^2_1 - m^2_w)^2}  \frac{q^2_2}{(q^2_2 - m^2_w)^2}
    16 \sqrt{X} \cdot \left\{ \rho^{(1)}_{00} \rho^{(2)}_{00}
    {\hat \sigma}_{LL} +  \right.
$$
\be
  + \left( \rho^{(1)}_{++}  \rho^{(2)}_{++} + \rho^{(1)}_{--} \rho^{(2)}_{--}
  \right) {\hat \sigma}_{TT} + 2 \left( \rho^{(1)}_{+0} \rho^{(2)}_{+0} +
  h.c. \right)  {\cal J}_{LT} +
\ee
$$
  \left. +
  \left( \rho^{(1)}_{+-} \rho^{(2)}_{+-} + h.c. \right) {\cal J}_{TT}
  \right\}  \frac{1}{2^8 \pi^4}  d \zeta d \eta d q^2_1 d q^2_2
  \frac{d (\Delta \phi)}{2 \pi} ~,
$$
where $\zeta = \frac{2w_1}{\sqrt S};~ \eta = \frac{2 w_2}{\sqrt S};~
\Delta \phi = \phi_1 - \phi_2$ -- is the relative azimuthal angle in
$e^+e^-$ -- c.m.s.
% 15
\be
    \cos \Delta \phi = - \frac{(q_{1 \perp} \cdot q_{2 \perp})}
    {\sqrt{q^2_{1 \perp} \cdot q^2_{2 \perp}}} ~,
\ee
where $q_{1 \perp} (q_{2 \perp})$ -- is the perpendicular momentum of
$W^+ (W^-)$ in $e^+e^-$ -- c.m.s., $\sqrt X$ -- is the f\/lux of the
virtual $W$--bosons:
% 16
\be
   X = (q_1 q_2)^2 - q^2_1 q^2_2  ~,
\ee
${\hat \sigma}_{LL}$, and ${\hat \sigma}_{TT}$ represent the "cross-section"
of the corresponding virtual $W$--boson components [6]:
% 17
\be
   {\hat \sigma}_{LL} = \frac{\pi^2 \alpha_w}{2 \sqrt X} m^2_w
   \frac{(q_1 q_2)^2}{q^2_1 q^2_2} ~\cdot~ 4 \delta
    \left( \hat s - M^2_H \right) ~,
\ee
%18
\be
   {\hat \sigma}_{TT} = \frac{\pi^2 \alpha_w}{2 \sqrt X} m^2_w
   ~\cdot~ 4 \delta  \left( \hat s - M^2_H \right)
\ee
and ${\cal J}_{LT}~, {\cal J}_{TT}$~ -- represent the interference
terms [6]:
% 19
\be
   {\cal J}_{LT} = \frac{\pi^2 \alpha_w}{2 \sqrt X} m^2_w
   \frac{(q_1 q_2)^2}{\sqrt{q^2_1 q^2_2}} ~\cdot~ 4 \delta
    \left( \hat s - M^2_H \right) ~,
\ee
%20
\be
   {\cal J}_{TT} = {\hat \sigma}_{TT} ~,
\ee
we have verif\/ied that exact calculation of the total cross-section (14)
with the substitution (15-20) gives exactly the same result as formula
(10) [12] (see Appendix IV).

Now, being sure of the normalization, we can look at the contribution of each
(17-20) term, multiplied by the corresponding density matrix element
in (14), into the total cross-section. The numerical
calculation of the contribution of the longitudinally and transversely
polarized
$W$--bosons into the total cross-section (14)
$\left( \sigma_{LL} + \sigma_{TT} \right)^{exact} /  \sigma^{exact}_{total}$
is plotted in Fig.2 for $M_H = 100 ~GeV$ as a function of $\sqrt s$.
One can see that it is systematically larger than the total cross-section.
This ef\/fect is easy to understand if we look at Fig.3, where the contribution
of each term in (14) is shown, along with the total cross-section. The main
contribution comes from the longitudinally polarized $W$--bosons, while the
contribution
of the transversely polarized $W$--bosons is small. Interestingly, the
spin-f\/lip term ${\cal J}_{LT}$, provides a large contribution with a negative
sign
which cancels a large part of the longitudinal contribution of $\sigma_{LL} $
into the $\sigma_{total} $.

The spin-f\/lip terms $\rho_{+0}$ and $\rho_{+-}$ in (14) depend on
$\Delta \tilde{\phi} $ -- the relative azimuthal angle def\/ined in the
$W^+W^-$ c.m.s. (9), which is dif\/ferent from $\Delta \phi$ (15)
in the phase space volume (14) (def\/ined in the $e^+e^-$ c.m.s.).

Would  $\Delta \phi$ coincide with $\Delta \tilde{\phi} $, the contribution
of interference terms into the total cross section was exactly zero.
But these two azymuthal angles coincide only if two c.m. systems moves along
their $Z$ axis. In general case $\Delta \tilde{\phi} $ depends on
$\Delta \phi$ (see Appendix IV) and the contribution of quantum mechanical
interference terms into cross section is non zero and rather large in the
region of not very high energy and small Higgs-boson mass ($M_H \approx
m_W$) [12,13]. This subtle point was missed in the literature.

{}From exact formula of Appendix III, taking
% 21
\be
   U = \frac{1}{2} \theta_1 \theta_2 (1 - \cos \Delta \varphi)
   +\frac12(\theta_1-\theta_2)^2
\ee
we estimate, that $e^+e^-$ lab. system and $WW$--c.m.s. becomes
acollinear when fermion scattering angles $\theta_{1,2}$ are greater
then:
%22
\be
   \theta_m > \frac{4m^2_w}{s \left( 1 - \frac{M^2_H}{s}
   \right) } ~.
\ee
The nonvanishing contribution from interference
terms ${\cal J}_{int}$  in the total cross-section, compared to the
longitudinal polarization contribution will be then:
% 23
\be
   \frac{{\cal J}_{LT}}{\sigma_{LL}} = - \frac{m^2_w}{M_H^2}
       ~~\frac{ln^2 \frac{s \theta_m}{4m^2_w} (1 - M^2_H /s)}
           {ln s/M^2_H} ~,
\ee
i.e. it vanishes both at small scattering angles, and when $M_H \gg m_w$.
\vspace{5mm}

%%%%%%%%%%%%%%%%%%%%%%%%%%%%%%%%%%%%%%%%%%%%%%%%%%%%%%%%%%%%%%%%%%%%%%%%

\underline{\large\bf  V. ~Ef\/fective vector--boson method}
\vspace{3mm}

%%%%%%%%%%%%%%%%%%%%%%%%%%%%%%%%%%%%%%%%%%%%%%%%%%%%%%%%%%%%%%%%%%%%%%%%%

The standard version of the ef\/fective $W$--boson method  [1, 2] could be
derived from (14) if one takes into account only the terms proportional
to ${\hat \sigma_{LL}}$ and ${\hat \sigma_{TT}}$.
Since (14) is already written in the
factorized form, it is easy to calculate the ef\/fective $W$-boson
distribution functions. Of course they coincide with the standard ones.
It is important now, to apply these functions only in the kinematical
region, where the new additional terms in (14) (which represents
quantum mechanical interference ef\/fects) are not important. From
Section 4, we have found that it can occur in the case of small
scattering angle of electrons (22). It means that there must be a cut
of\/f in the integral over $d q^2_i$, as $q^2_i$ is connected with
$\cos \theta_i$:
% 24
\be
   q^2_1 = - \frac{s}{2} (1 - \cos \theta_1) \cdot (1 - \zeta) ~.
\ee

We have performed computer calculations for each helicity component in
the total cross-section $\sigma^{tot}$ and for $\sigma^{tot}$  itself
for dif\/ferent value of $q^2$. The result is presented in Fig.4.
We see that the transversal cross section $\sigma_{TT}$  is small
compared to $\sigma^{tot}$ for any $q^2$, and its contribution can be
neglected. In the region $q^2 \ge M^2_H = 100~ GeV^2$ interference term
--$\tau_{LT} (q^2)$ coincides with $\sigma_{LL}$ and the sum
$(\sigma_{LL} + \sigma_{TT})(q^2)$ doesn't correspond to the
$\sigma^{tot}$  for this value of $q^2$ as it was always considered in
the previous calculations. So the natural cut-off for $q^2$
is $\Lambda^2 = \hat s$, where $\hat s$ is a characteristic energy of
$W^+ W^-$ cross-section. The same restriction on $q^2$ is needed
to have $W$-boson beams to be independent from each other [6].
(For ef\/fective photons this cut-of\/f was considered much earlier in
[7].)

For $q^2 < \hat s = M^2_H$ we have the following approximate formula for
$\sigma_{LL}$ and  $\sigma_{TT}$:
% 25
\be
   \sigma^{appr}_{LL} = \left( \frac{\alpha}{\sin^2 \theta_w} \right)^3
     \frac{1}{16m^2_w}  \frac{1}{ \left(  1 +
      \frac{m^2_w}{\Lambda^2} \right)^2 } \left[ \left( 1 +
      \frac{M^2_H}{s} \right) ln \frac{s}{M^2_H} - 2 \left( 1 -
      \frac{M^2_H}{s} \right) \right] ~,
\ee
\vspace{3mm}
% 26
$$
   \sigma^{appr}_{TT} = \left( \frac{\alpha}{\sin^2 \theta_w} \right)^3
     \frac{1}{16m^2_w}  \frac{m^4_w}{M_H}
       \left[ 2 \left( 1 + \frac{2M^2_H}{s} + \frac{M^4_H}{2s^2}  \right)
      ln \frac{s}{M^2_H} - 3 \left( 1 - \frac{M^4_H}{s^2} \right) \right]
       \cdot
$$
\be
   \cdot \left[ ln \left( \frac{\Lambda^2}{m^2_w} + 1 \right)
    - \frac{1}{1 + \frac{M^2_H}{\Lambda^2}}  \right]^2 ~.
\ee

If we'll absolutely neglect ($-\tau_{LT}$) in this region $q^2 < \Lambda^2$
then we can calculate approximate total cross section. The ratios of
$\sigma^{appr}_{LL}/ \sigma_{total}^{exact}$ and $(\sigma_{LL} +
\sigma_{TT} )^{appr} / \sigma^{exact}_{total}$ as functions of $\sqrt
s$ are plotted in Fig.5(a,b). We have taken cut-of\/f exactly
$\Lambda^2 = M^2_H)$ and reproduce $\sigma^{exact}_{total}$ with
rather good accuracy.  In principle one can also analytically take
into account the small contribution of interference term
$-\tau_{LT}$ (23). In this case we expect to obtain a little bit
smaller cross section just because $\sigma^{appr}_{tot}$ is obtained
from a small part of the phase space, in the same way as we already
know from $\gamma\gamma$--physics [7, 14].

Note, that the same formulas (25-26) are well applied in the case of high
energy and heavy Higgs $M_H \gg m_w$ with $\Lambda^2 = M^2_H$, and give the
same result as in [2, 3, 10, 11] with a strong suppression of the
transversely polarizeed $W$-beams, as
% 27
\be
   \frac{\sigma_{TT}}{\sigma_{LL}} ~\sim~ \left( \frac{m_w}{M_H}
        \right)^4 .
\ee
\vspace{5mm}

%%%%%%%%%%%%%%%%%%%%%%%%%%%%%%%%%%%%%%%%%%%%%%%%%%%%%%%%%%%%%%%%%%%%%%%%%%

\underline{\large\bf  C o n c l u s i o n s}
\vspace{3mm}

%%%%%%%%%%%%%%%%%%%%%%%%%%%%%%%%%%%%%%%%%%%%%%%%%%%%%%%%%%%%%%%%%%%%%%%%%

We have demonstrated that the approximate ef\/fective $W$-boson method
can be well applied in the kinematical region where the $W^+W^-$
c.m.s. is collinear with the c.m.s. of the colliding particles, i.e.
in the case of small scattering angles for the electrons (or quarks)
and gives the lower bound of the cross-section. It means that there must
be introduced a cut-of\/f $\Lambda^2 \lsim M^2_H$ when $M_H \gg m_w)$
or $\Lambda^2 \lsim m^2_w$ (when $M_H \approx m_w$) in the integrals over
the $q^2-W$-boson virtualities. In this region the contribution of the
transversely polarized beams of $W$-bosons are always smaller than
the leading contribution of longitudinally polarized $W$-beams.

It was shown, that when these two c.m.s. are not the collinear ones, the
negative sign contribution of interference terms plays a crucial role and must
be taken into account. That fact was missed in the previous works on this
subject.

It is important that the nondiagonal structure of $W$-boson density matrix
is def\/ined only by the fermion-$W$-fermion vertex and does not depend
on the details of $WW$-interaction. But of course the contribution of each term
of the density matrix in the particular subprocess will depend on the
dynamics of the $WW$-interaction and its particular kinematics. So in each
particular case all the terms (not only the transversely and longitudinally
polarized ones) of the density matrix must be taken into consideration.
\vspace{5mm}

%%%%%%%%%%%%%%%%%%%%%%%%%%%%%%%%%%%%%%%%%%%%%%%%%%%%%%%%%%%%%%%%%%%%%%%%%%

\underline{\large\bf  Acknowledgements}
\vspace{3mm}

%%%%%%%%%%%%%%%%%%%%%%%%%%%%%%%%%%%%%%%%%%%%%%%%%%%%%%%%%%%%%%%%%%%%%%%%%

We are greatful to Ph.Bambade for the help with the numerical calculations.
A. D. would like to thank P.Kessler, J.Parisi and S.Ong for stimulating
discussions on the analogy with QED and equivalent photon approximation
method and acknowledge useful discussions  with A.B.Kaidalov and
M.Ciafaloni.

\newpage
%%%%%%%%%%%%%%%%%%%%%%%%%%%%%%%%%%%%%%%%%%%%%%%%%%%%%%%%%%%%%%%%%%%%%%%%%%

\hfill  \underline{\bf Appendix I}
\vspace{8mm}
%%%%%%%%%%%%%%%%%%%%%%%%%%%%%%%%%%%%%%%%%%%%%%%%%%%%%%%%%%%%%%%%%%%%%%%%%%%%

The virtual $W$-boson density matrix elements:
\vspace{5mm}

\begin{tabular}{lcl}
  $ \rho^{(1)}_{00}$ & = & $\frac{s^2}{X} \left[ \left( \eta -
   \frac{q^2_2}{s} - \frac{(q_1q_2)}{s} \right)^2 \frac{X}{s^2}
     \right]$ ~, \\  [3mm]
$\rho^{(1)}_{\pm \pm}$ & = & $\frac{s^2}{2X} \left[ \frac{2X}{s^2} -
\left( \eta - \frac{q^2_2}{s} \right) \left( \frac{\hat s}{s} -
     \frac{q^2_1}{s} - \eta \right) + \frac{q^2_1 q^2_2}{s^2}  \pm
      \frac{2\sqrt X}{s} \left( \eta - \frac{q^2_2}{s} - \frac{(q_1 q_2)}{s}
       \right) \right]$ ~, \\  [3mm]
$\rho^{(1)}_{+-}$ & = & $\frac{e^{2i \tilde{\varphi}}}{2} \frac{s^2}{X}
     \left[ \left( \frac{\hat s}{s} - \frac{q^2_2}{s} - \eta \right)
     \left( \eta - \frac{q^2_2}{s} \right) - \frac{q^2_1 q^2_2}{s^2}
     \right]$ ~, \\  [3mm]
$\rho^{(1)}_{\stackrel{+ 0}{0 -}}$  & =
     & $- \frac{i e^{i \tilde{\varphi}}}{\sqrt 2}
    \frac{s^2}{X} \left\{ -\eta + \frac{q^2_2}{s} + \frac{(q_1 q_2)}{s}
     \pm \frac{\sqrt X}{s} \right\} \cdot \left[ \frac{q^2_1 q^2_2}{s^2}
      - \left( \eta - \frac{q^2_2}{s} \right) \left( \frac{\hat s}{s}
        - \frac{q^2_2}{s} - \eta \right) \right]^{1/2}$ ~.
\end{tabular}
\vspace{25mm}

%%%%%%%%%%%%%%%%%%%%%%%%%%%%%%%%%%%%%%%%%%%%%%%%%%%%%%%%%%%%%%%%%%%%%%%%%%

\hfill  \underline{\bf Appendix II}
\vspace{8mm}
%%%%%%%%%%%%%%%%%%%%%%%%%%%%%%%%%%%%%%%%%%%%%%%%%%%%%%%%%%%%%%%%%%%%%%%%%%%%

The virtual $W$-boson density matrix in the case of small scattering
angles of the colliding fermions [10]:

$$
\rho^{(1)}_{00}  ~=  \frac{4}{\zeta^2} (1 - \zeta) ~,~~~~~~~~~~~~~~~~~~~~~
$$
$$
\rho^{(1)}_{++} = \frac{2}{\zeta^2} ~,~~~~~~~~~~~~~~~~~~~~~~~~~~~~~~
$$
$$
\rho^{(1)}_{--} = \frac{2}{\zeta^2}  (1 - \zeta) ~,~~~~~~~~~~~~~~~~~~~~~
$$
$$
\rho^{(1)}_{+0} ~= - 2 \sqrt{2} i e^{i \tilde{\varphi}_1}
                    \frac{1}{\zeta^2} (1 - \zeta)^{1/2} ~,~~~~
$$
$$
\rho^{(1)}_{0-} ~= - 2 \sqrt{2} i e^{i \tilde{\varphi}_1}
                    \frac{1}{\zeta^2} (1 - \zeta)^{3/2} ~,~~~~~
$$
$$
\rho^{(1)}_{+-} = - 2  e^{2i \tilde{\varphi}_1}
                    \frac{1}{\zeta^2} (1 - \zeta) ~,~~~~~~~~~~~
$$
where $\zeta = \frac{2 \omega_1}{\sqrt s}$. The analogous relation for
$\rho^{(2)}_{ab}$ can be found with a substitution
$\frac{2 \omega_1}{\sqrt s} \to \frac{2 \omega_2}{\sqrt s} = \eta;
{}~\varphi_1 \to -\varphi_2$ (and $q^2_2 \to q^2_1$ in Appendix I).

\newpage
%%%%%%%%%%%%%%%%%%%%%%%%%%%%%%%%%%%%%%%%%%%%%%%%%%%%%%%%%%%%%%%%%%%%%%%%%%

\hfill  \underline{\bf Appendix III}
\vspace{8mm}
%%%%%%%%%%%%%%%%%%%%%%%%%%%%%%%%%%%%%%%%%%%%%%%%%%%%%%%%%%%%%%%%%%%%%%%%%%%%

For the exact numerical calculation of the total cross-section (10-13),
we used the following f\/inal formula:
$$
  \frac{d \sigma \cdot 2 \pi}{d \Delta \varphi} = \frac{\alpha^3_w m^2_w}
    {4s^2} \int^1_{-1} d \cos \theta_1  \int^1_{-1} d \cos \theta_2
       \int^1_{M^2_H/s}   d \zeta  ~\cdot
$$
$$
   \cdot \frac{1}{\left( 1 + \frac{2m^2_w}{s(1-\eta_0)} - \cos \theta_2
      \right)^2 \cdot \left(1 + \frac{2 m^2_w}{s(1 - \zeta)} - \cos \theta_1
        \right)^2 } \cdot \frac{(1- U)}{U + (1 - U) \zeta} ~;
$$
where $\eta_0 = \frac{m^2_H/s + (1-\zeta) U}{\zeta (1 - U) + U}; ~U =
\frac{1}{2} \left( 1 - \cos \theta_1 \cdot \cos \theta_2 - \sin \theta_1
\sin \theta_2 \cos \Delta \varphi \right)$
$$
  ( 0 ~\le~ U ~\le~ 1 ).
$$
\vspace{15mm}

%%%%%%%%%%%%%%%%%%%%%%%%%%%%%%%%%%%%%%%%%%%%%%%%%%%%%%%%%%%%%%%%%%%%%%%%%%

\hfill  \underline{\bf Appendix IV}
\vspace{3mm}
%%%%%%%%%%%%%%%%%%%%%%%%%%%%%%%%%%%%%%%%%%%%%%%%%%%%%%%%%%%%%%%%%%%%%%%%%%%%

For the exact numerical calculations of the helicity components in the total
cross-section (14) (with (9), (15)-(20) ) ~we used as the f\/inal formula:
$$
  \frac{2 \pi d \sigma}{d \Delta \varphi} = \frac{m^2_w \alpha^3_w}
    {2^6 s^2} \int^1_{-1}  d \cos \theta_1  \int^1_{-1} d \cos \theta_2
      \int^1_{\frac{M^2_H}{S}}   d \zeta  ~\cdot
$$
$$
  \cdot
  \left\{ \rho^{(1)}_{00} \rho^{(2)}_{00} \frac{(q_1 q_2)^2}{q^2_1
  q^2_2} + \rho^{(1)}_{++} \rho^{(2)}_{++}  + \rho^{(1)}_{--}
     \rho^{(2)}_{--} + 2 \cos 2 \Delta \tilde{\varphi} |
     \rho^{(1)}_{+-} |~ | \rho^{(2)}_{+-} | ~-
   \right.
$$
$$
\left.
   -~ \frac{2(q_1 q_2)}{\sqrt{q^2_1 q^2_2}}  \cos \Delta \tilde{\varphi}
       \left(  | \rho^{(1)}_{+0} |~ | \rho^{(2)}_{+0} | +  |
       \rho^{(1)}_{0 -} |~ | \rho^{(2)}_{0 -} | \right)  \right\} ~,
$$
where $\eta_0, ~U$ are the same as in the Appendix III;
$$
  q^2_2 = - \frac{s}{2}(1 - \eta_0)(1 - \cos \theta_2);
  ~q^2_1 = - \frac{s}{2}(1 - \zeta)(1 - \cos \theta_1);
  ~(q_1q_2) = \frac{M^2_H - q^2_1 - q^2_2}{2};
$$
$$
  \cos \Delta \tilde{\varphi} = \frac{s}{2 \sqrt{q^2_1 q^2_2}}  ~\cdot
$$
$$
   \frac{\eta_0 \zeta \left( \frac{M^2_H}{s} - \frac{q^2_1}{s} -
\frac{q^2_2}{s}
    \right) - \zeta \frac{q^2_2}{s} \left( \frac{M^2_H}{s} + \frac{q^2_1}{s}
     - \frac{q^2_2}{s} \right) -  \eta_0 \frac{q^2_1}{s}
      \left( \frac{M^2_H}{s} + \frac{q^2_2}{s}- \frac{q^2_1}{s} \right) -
       \frac{4X}{s^2} + \frac{2M^2_H q^2_1 q^2_2}{s^3}  }
   { \left[ \left( \zeta - \frac{q^2_1}{s} \right)
            \left( \zeta - \frac{M^2_H}{s} \right)  + \zeta \cdot
       \frac{q^2_2}{s}  \right]^{1/2}  \left[ \left( \eta_0 -
        \frac{q^2_2}{s} \right) \left( \eta_0 - \frac{M^2_H}{s} \right)
     + \eta_0 \frac{q^2_1}{s} \right]^{1/2} }
$$
$$
  \cos 2 \Delta \tilde{\varphi} = 2 \cos^2  \Delta \tilde{\varphi} - 1 ~.
$$

\newpage

\newpage

%%%%%%%%%%%%%%%%%%%%%%%%%%%%%%%%%%%%%%%%%%%%%%%%%%%%%%%%%%%%%%%%%%%%%%%%%%%%

\centerline{\underline\bf FIGURE CAPTIONS}
\vspace{15mm}

\parbox[t]{2cm}{Fig. 1}
\parbox[t]{130mm}
                {Feynman graph for Higgs-boson production via $W^+W^-$
                fusion in $e^+e^-$ (or $q \bar q$) collisions.  }
\vspace{3mm}

\parbox[t]{2cm}{Fig. 2}
\parbox[t]{130mm}
               {The exact numerical calculation of the contribution of only
               diagonal terms of $W$-density matrix into the total
               cross-section. It is seen that without interference
               terms it overstimates almost 3 times the total
               cross-section.  }
\vspace{3mm}

\parbox[t]{2cm}{Fig. 3}
\parbox[t]{130mm}
                {The exact numerical calculation of the contribution of each
                polarization of the $W$-amplitude in the total
                cross-section. It is very well seen the cancelation
                of $\sigma_{LL}$ by $\cal J_{\rm LT}$ - the interference term.}
\vspace{3mm}

\parbox[t]{2cm}{Fig. 4}
\parbox[t]{130mm}
               {The exact computer calculation of the dependence of each
               polarization of $W$-amplitude on the $W$-boson
               virtuality $q^2$. One can see that at $q^2 = \Lambda^2 =
               M^2_H$ there is the total cancelation between longitudinal
               and interference terms. It means that the integration over
               $d q^2$ in the approximate calculations via ef\/fective
               $W$-boson method has a cut-of\/f $\Lambda \sim M_H$. }
\vspace{3mm}

\parbox[t]{2cm}{Fig. 5}
\parbox[t]{130mm}
              {The ratio of the approximate ef\/fective $W$-boson method
               calculation of $\sigma^{appr}_{LL}$ (a) and
               $(\sigma_{LL}^{ww} + \sigma_{TT}^{ww})^{appr}$  (b)  with
               $q^2 < \Lambda^2 = M^2_H$ over the exact computer
               calculation $\sigma_{exact}$. The ratio is always less
               than 1 above the threshold, it means that the ef\/fective
               $W$-boson method gives the lower bound of the cross
               section.  }
\end{document}